\documentclass[10pt,twocolumn]{IEEEtran}
\usepackage{graphicx,subfigure}
\usepackage{psfig,epsfig,algorithm,amsmath,amssymb,color,subfigure,url}
\usepackage[center]{caption}

\begin{document}
\setlength{\arraycolsep}{0.03cm}
\newcommand{\xhat}{\hat{x}}
\newcommand{\xpred}{\hat{x}_{t|t-1}}
\newcommand{\xupd}{\hat{x}_{t|t}}
\newcommand{\Ppred}{P_{t|t-1}}
\newcommand{\ty}{\tilde{y}_t}
\newcommand{\tty}{\tilde{y}_{t,\text{res}}}
\newcommand{\tw}{\tilde{w}_t}
\newcommand{\ttw}{\tilde{w}_{t,f}}
\newcommand{\betahat}{\hat{\beta}}

\newcommand{\ypast}{y_{1:t-1}}
\newcommand{\sone}{S_{*}}
\newcommand{\sinf}{{S_{**}}}
\newcommand{\smax}{S_{\max}}
\newcommand{\smin}{S_{\min}}
\newcommand{\samax}{S_{a,\max}}
\newcommand{\Nhat}{{\hat{N}}}

\newcommand{\Dnum}{D_{num}}
\newcommand{\pss}{p^{**,i}}
\newcommand{\fr}{f_{r}^i}

\newcommand{\A}{{\cal A}}
\newcommand{\Z}{{\cal Z}}
\newcommand{\B}{{\cal B}}
\newcommand{\R}{{\cal R}}
\newcommand{\reg}{{\cal G}}
\newcommand{\const}{\mbox{const}}

\newcommand{\trace}{\mbox{tr}}

\newcommand{\hsim}{{\hspace{0.0cm} \sim  \hspace{0.0cm}}}
\newcommand{\he}{{\hspace{0.0cm} =  \hspace{0.0cm}}}

\newcommand{\vect}[2]{\left[\begin{array}{cccccc}
     #1 \\
     #2
   \end{array}
  \right]
  }

\newcommand{\matr}[2]{ \left[\begin{array}{cc}
     #1 \\
     #2
   \end{array}
  \right]
  }
\newcommand{\vc}[2]{\left[\begin{array}{c}
     #1 \\
     #2
   \end{array}
  \right]
  }

\newcommand{\gdot}{\dot{g}}
\newcommand{\Cdot}{\dot{C}}
\newcommand{\re}{\mathbb{R}}
\newcommand{\n}{{\cal N}}  
\newcommand{\N}{{\overrightarrow{\bf N}}}  
\newcommand{\chat}{\tilde{C}_t}
\newcommand{\chati}{\chat^i}

\newcommand{\cmin}{C^*_{min}}
\newcommand{\twi}{\tilde{w}_t^{(i)}}
\newcommand{\twj}{\tilde{w}_t^{(j)}}
\newcommand{\wi}{{w}_t^{(i)}}
\newcommand{\twio}{\tilde{w}_{t-1}^{(i)}}

\newcommand{\tWi}{\tilde{W}_n^{(m)}}
\newcommand{\tWj}{\tilde{W}_n^{(k)}}
\newcommand{\Wi}{{W}_n^{(m)}}
\newcommand{\tWio}{\tilde{W}_{n-1}^{(m)}}

\newcommand{\ds}{\displaystyle}

\newcommand{\SAR}{S$\!$A$\!$R }
\newcommand{\MAR}{MAR}
\newcommand{\MMRF}{MMRF}
\newcommand{\AR}{A$\!$R }
\newcommand{\GMRF}{G$\!$M$\!$R$\!$F }
\newcommand{\DTM}{D$\!$T$\!$M }
\newcommand{\MSE}{M$\!$S$\!$E }
\newcommand{\RCS}{R$\!$C$\!$S }
\newcommand{\uomega}{\underline{\omega}}
\newcommand{\y}{v}
\newcommand{\x}{w}
\newcommand{\lu}{\mu}
\newcommand{\g}{g}
\newcommand{\s}{{\bf s}}
\newcommand{\bft}{{\bf t}}
\newcommand{\refmap}{{\cal R}}
\newcommand{\totrefl}{{\cal E}}
\newcommand{\beq}{\begin{equation}}
\newcommand{\eeq}{\end{equation}}
\newcommand{\bdm}{\begin{displaymath}}
\newcommand{\edm}{\end{displaymath}}
\newcommand{\hatz}{\hat{z}}
\newcommand{\hatu}{\hat{u}}
\newcommand{\tilz}{\tilde{z}}
\newcommand{\tilu}{\tilde{u}}
\newcommand{\hhatz}{\hat{\hat{z}}}
\newcommand{\hhatu}{\hat{\hat{u}}}
\newcommand{\tilc}{\tilde{C}}
\newcommand{\hatc}{\hat{C}}
\newcommand{\tim}{n}

\newcommand{\ssp}{\renewcommand{\baselinestretch}{1.0}}
\newcommand{\defd}{\mbox{$\stackrel{\mbox{$\triangle$}}{=}$}}
\newcommand{\goes}{\rightarrow}
\newcommand{\tends}{\rightarrow}
\newcommand{\defn}{\triangleq} 
\newcommand{\se}{&=&}
\newcommand{\sdefn}{& \defn  &}
\newcommand{\sle}{& \le &}
\newcommand{\sge}{& \ge &}
\newcommand{\plusminus}{\stackrel{+}{-}}
\newcommand{\Ey}{E_{Y_{1:t}}}
\newcommand{\ey}{E_{Y_{1:t}}}

\newcommand{\equivto}{\mbox{~~~which is equivalent to~~~}}
\newcommand{\nonzero}{i:\pi^n(x^{(i)})>0}
\newcommand{\nonzeroc}{i:c(x^{(i)})>0}

\newcommand{\supn}{\sup_{\phi:||\phi||_\infty \le 1}}
\newtheorem{remark}{Remark}
\newtheorem{example}{Example}
\newtheorem{ass}{Assumption}
\newtheorem{proposition}{Proposition}

\newtheorem{fact}{Fact}
\newtheorem{heuristic}{Heuristic}
\newcommand{\eps}{\epsilon}
\newcommand{\bd}{\begin{definition}}
\newcommand{\ed}{\end{definition}}
\newcommand{\udq}{\underline{D_Q}}
\newcommand{\td}{\tilde{D}}
\newcommand{\epsinv}{\epsilon_{inv}}
\newcommand{\al}{\mathcal{A}}

\newcommand{\bfx} {\bf X}
\newcommand{\bfy} {\bf Y}
\newcommand{\bfz} {\bf Z}
\newcommand{\ddas}{\mbox{${d_1}^2({\bf X})$}}
\newcommand{\ddbs}{\mbox{${d_2}^2({\bfx})$}}
\newcommand{\dda}{\mbox{$d_1(\bfx)$}}
\newcommand{\ddb}{\mbox{$d_2(\bfx)$}}
\newcommand{\xinc}{{\bfx} \in \mbox{$C_1$}}
\newcommand{\eqa}{\stackrel{(a)}{=}}
\newcommand{\eqb}{\stackrel{(b)}{=}}
\newcommand{\eqe}{\stackrel{(e)}{=}}
\newcommand{\leqc}{\stackrel{(c)}{\le}}
\newcommand{\leqd}{\stackrel{(d)}{\le}}

\newcommand{\leqa}{\stackrel{(a)}{\le}}
\newcommand{\leqb}{\stackrel{(b)}{\le}}
\newcommand{\leqe}{\stackrel{(e)}{\le}}
\newcommand{\leqf}{\stackrel{(f)}{\le}}
\newcommand{\leqg}{\stackrel{(g)}{\le}}
\newcommand{\leqh}{\stackrel{(h)}{\le}}
\newcommand{\leqi}{\stackrel{(i)}{\le}}
\newcommand{\leqj}{\stackrel{(j)}{\le}}

\newcommand{\w}{{W^{LDA}}}
\newcommand{\halpha}{\hat{\alpha}}
\newcommand{\hsigma}{\hat{\sigma}}
\newcommand{\slmax}{\sqrt{\lambda_{max}}}
\newcommand{\slmin}{\sqrt{\lambda_{min}}}
\newcommand{\lmax}{\lambda_{max}}
\newcommand{\lmin}{\lambda_{min}}

\newcommand{\da} {\frac{\alpha}{\sigma}}
\newcommand{\chka} {\frac{\check{\alpha}}{\check{\sigma}}}
\newcommand{\sumo}{\sum _{\underline{\omega} \in \Omega}}
\newcommand{\distance}{d\{(\hatz _x, \hatz _y),(\tilz _x, \tilz _y)\}}
\newcommand{\col}{{\rm col}}
\newcommand{\rcs}{\sigma_0}
\newcommand{\CalR}{{\cal R}}
\newcommand{\df}{{\delta p}}
\newcommand{\dq}{{\delta q}}
\newcommand{\dZ}{{\delta Z}}
\newcommand{\pprime}{{\prime\prime}}

\newcommand{\vn}{N}

\newcommand{\bv}{\begin{vugraph}}
\newcommand{\ev}{\end{vugraph}}
\newcommand{\bi}{\begin{itemize}}
\newcommand{\ei}{\end{itemize}}
\newcommand{\ben}{\begin{enumerate}}
\newcommand{\een}{\end{enumerate}}
\newcommand{\be}{\protect\[}
\newcommand{\ee}{\protect\]}
\newcommand{\bean}{\begin{eqnarray*} }
\newcommand{\eean}{\end{eqnarray*} }
\newcommand{\bea}{\begin{eqnarray} }
\newcommand{\eea}{\end{eqnarray} }
\newcommand{\nn}{\nonumber}
\newcommand{\ba}{\begin{array} }
\newcommand{\ea}{\end{array} }
\newcommand{\ep}{\mbox{\boldmath $\epsilon$}}
\newcommand{\epp}{\mbox{\boldmath $\epsilon '$}}
\newcommand{\Lep}{\mbox{\LARGE $\epsilon_2$}}
\newcommand{\und}{\underline}
\newcommand{\pdif}[2]{\frac{\partial #1}{\partial #2}}
\newcommand{\odif}[2]{\frac{d #1}{d #2}}
\newcommand{\dt}[1]{\pdif{#1}{t}}
\newcommand{\urho}{\underline{\rho}}

\newcommand{\spc}{{\cal S}}
\newcommand{\tspc}{{\cal TS}}

\newcommand{\uv}{\underline{v}}
\newcommand{\us}{\underline{s}}
\newcommand{\uc}{\underline{c}}
\newcommand{\utheta}{\underline{\theta}^*}
\newcommand{\ualpha}{\underline{\alpha^*}}

\newcommand{\uxy}{\underline{x}^*}
\newcommand{\uxyj}{[x^{*}_j,y^{*}_j]}
\newcommand{\arcl}[1]{arclen(#1)}
\newcommand{\one}{{\mathbf{1}}}

\newcommand{\uxyjt}{\uxy_{j,t}}
\newcommand{\E}{\mathbb{E}}

\newcommand{\rhomat}{\left[\begin{array}{c}
                        \rho_3 \ \rho_4 \\
                        \rho_5 \ \rho_6
                        \end{array}
                   \right]}
\newcommand{\deltat}{\tau} 
\newcommand{\deltatt}{\Delta t_1}
\newcommand{\ceil}[1]{\ulcorner #1 \urcorner}

\newcommand{\xxi}{x^{(i)}}
\newcommand{\txi}{\tilde{x}^{(i)}}
\newcommand{\txj}{\tilde{x}^{(j)}}

\newcommand{\mi}[1]{{#1}^{(m,i)}}

\newcommand{\Section}[1]{ \vspace{-0.05in}  \section{#1}  \vspace{-0.05in} }
\newcommand{\Subsection}[1]{ \vspace{-0.1in}  \subsection{#1} \vspace{-0.05in}  }
\title{Support-Predicted Modified-CS for Recursive Robust Principal Components' Pursuit}
\author{
\authorblockN{Chenlu Qiu and Namrata Vaswani}\\
\authorblockA{Dept. of Electrical and Computer Engineering, Iowa State University, Ames, IA 50011, USA \\ Email: chenlu, namrata@iastate.edu
}
}
\maketitle

\begin{abstract}

This work proposes a causal and recursive algorithm for solving the ``robust" principal components' analysis (PCA) problem. We primarily focus on robustness to correlated outliers. 
In recent work, we proposed a new way to look at this problem and showed how a key part of its solution strategy involves solving a noisy compressive sensing (CS) problem. However, if the support size of the outliers becomes too large, for a given dimension of the current PC space, then the number of ``measurements" available for CS may become too small.
In this work, we show how to address this issue by utilizing the correlation of the outliers to predict their support at the current time; and using this as ``partial support knowledge" for solving Modified-CS instead of CS.

\end{abstract}

\Section{Introduction}

In this work, we propose a real-time (causal and recursive) algorithm for solving the ``robust" principal components' analysis (PCA) problem. Here, ``robust" refers to robustness to both independent and correlated sparse outliers, although we focus on the latter. The goal of PCA is to find the principal components' (PC) space, which is the smallest dimensional subspace that spans (or, in practice, approximately spans) a given dataset. Computing the PCs in the presence of outliers is called robust PCA. If the PC space changes over time, there is a need to update the PCs.
Doing this recursively is referred to as ``recursive robust PCA". Often the data vectors (both the low rank component and the outliers) are correlated.

A key application where this problem occurs is in automatic video surveillance where the goal is to separate a slowly changing background from moving foreground objects. The background sequence is well modeled by a low rank subspace that can gradually change over time, while the moving foreground objects constitute the ``correlated sparse outliers". \emph{We will use this as the motivating problem in this entire paper.} Other important applications include  sensor networks based detection and tracking of abnormal events such as forest fires or oil spills; or online detection of brain activation patterns from fMRI sequences (the ``active" part of the brain can be interpreted as a correlated sparse outlier).

Note that even though \emph{we use the term ``outlier" for the moving objects or the brain active regions etc, quite often, these are actually the ``signals of interest"}. But for PCA (used to estimate the low rank subspace of the background), they are the ``outliers" that make the PCA problem difficult.

The recursive robust PCA problem, that we study in this work, can be precisely defined as follows.
The outlier-corrupted data vectors (we will call them measurements), $M_t$, at each time $t$, can be rewritten as
\begin{equation}
M_t = L_t  + S_t \label{eq1}
\end{equation}
where $L_t$ is the ``low-rank" part and $S_t$ is the sparse outlier which can be correlated over time. We put ``low-rank" in quotes since it does not mean anything for a single vector. To make this precise, we can rewrite
\begin{equation*}
L_t = U x_t
\end{equation*}
where $U$ is an unknown orthonormal matrix and $x_t$ is a sparse vector whose support changes slowly over time and whose elements are spatially uncorrelated. Let $N_t$ denote the support of $x_t$, and assume that it is piecewise constant with time. Then the columns of the sub-matrix, $P_t := (U)_{N_t}$ span the low rank subspace in which the current set of $L_t$'s lie and $L_t = P_t (x_t)_{N_t}$. We refer to $P_t$ as the principal components (PC) matrix. Clearly, this is also piecewise constant with time. If at a given time, $t=t_0$, $N_t$ changes, then $P_t$ also changes. New directions get added to it and some directions get removed from it.
We assume realistic correlation models on both $x_t$ (and hence $L_t$) and on $S_t$. These are defined in Sec. \ref{models}.

Suppose, we have a good estimate of the initial PC matrix\footnote{In most applications, it is easy to get a short sequence with no moving objects that can serve as the ``training data" to estimate $\hat{P}_0$ by simple PCA.}, $\hat{P}_0 \approx P_0$. For $t>0$, our goal is to recursively keep estimating the sparse part, $S_t$, at each time, and to keep updating $\hat{P}_t$ every-so-often, by collecting the recent estimates of $L_t=M_t-S_t$.

In (\ref{eq1}), if $U$ were known, and if we did not want a recursive solution, then this problem would be similar to the dense error correction  problem studied in recent work \cite{error_correction_PCP_l1, Laska_exactsignal}. Of course the reason we need to do PCA is because $U$ is \emph{unknown}.

There has been a large amount of earlier work on 
robust PCA e.g. \cite{rpca_2,rpca_xu} and recursive robust PCA e.g. \cite{sequentialSVD,ipca_weightedand}. In these works, (i) either the locations of the missing/corruped data points are assumed known \cite{sequentialSVD}, which is not a practical assumption, or (ii) they first try to detect the corrupted pixels and then either fill in the corrupted location using some heuristics or (iii) often just remove the entire outlier vector \cite{rpca_2,rpca_xu,ipca_weightedand}.
Very often, outliers affect only a small part of the data vector, e.g., the moving objects may occupy only a small image region. In a series of recent works \cite{rpca,rpca_Chandrasekaran,error_correction_PCP}, an elegant solution has been proposed, that treats the outliers as sparse vectors and hence does not require a two step outlier location detection/correction process and also does not throw out the entire vector. In \cite{rpca,rpca_Chandrasekaran,error_correction_PCP}, the offline ``robust PCA" problem is redefined as a problem of separating a low rank matrix, $L := [L_1,\dots,L_t]$, from a sparse matrix, $S := [S_1,\dots,S_t]$, using the data matrix, $M := [M_1,\dots,M_t] = L + S$. It was shown that by solving
$\min \|L\|_* + \lambda \|S\|_1, \ s.t. \ M = L+S$
(where $\|L\|_*$ is the sum of singular values of $L$ while $\|S\|_1$ is the $\ell_1$ norm of $S$ seen as a long vector), one can recover $L$ and $S$ exactly provided the singular vectors of $L$ are spread out enough (not sparse), the support and signs of $S$ are uniformly random (thus it is not low rank) and the rank of $L$ is sufficiently small for a given sparsity of $S$. This was called Principal Components' Pursuit (PCP).
PCP was motivated as a tool for video surveillance applications to track moving objects (sparse part) from the slow changing background (low rank part).

\begin{figure}
\centerline{
\psfig{file = 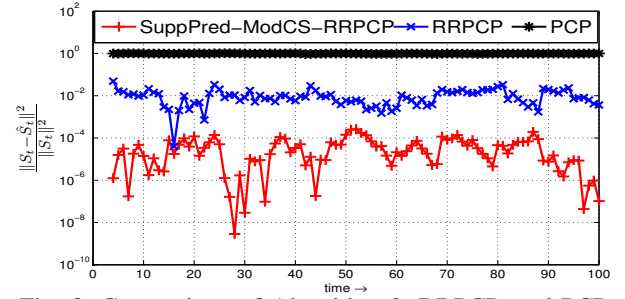, width = 8cm, height = 4cm}
}
\vspace{-0.1in}
\caption{Comparison of Algorithm \ref{algo2}, RRPCP, and PCP}
\vspace{-0.15in}
\label{plot1}
\end{figure}

While PCP is an elegant idea, it has three practical limitations.
In surveillance applications, one would like to obtain the estimates on-the-fly and quickly as a new frame comes in, rather than in a batch fashion. Second, in many applications, the support sets over time will be heavily correlated, and often, also overlapping, e.g. in the case of moving objects forming the sparse part. This can result in $S$ being low rank and thus makes it impossible for PCP to separate $S$ from $L$ [see Fig. \ref{plot1}]. Third, PCP requires the support of $x_t$ to be fixed and quite small for a given support size of $S_t$, e.g. see Table 1 of \cite{rpca}. But, often, this does not hold.
%
%
%

To address the first two drawbacks, in \cite{rrpcp_allerton}, we proposed a recursive robust PCP (RRPCP) algorithm that was also ``robust" to time-correlated sparse outlier sequences. In this work, \emph{we show how we can use the time-correlation of the ``outliers" to our advantage to address the third limitation above.} 

\Subsection{Motivation and Key Ideas}

\begin{figure*}[t!]
\centerline{
\subfigure[Recursive Robust PCP]{
\psfig{file = 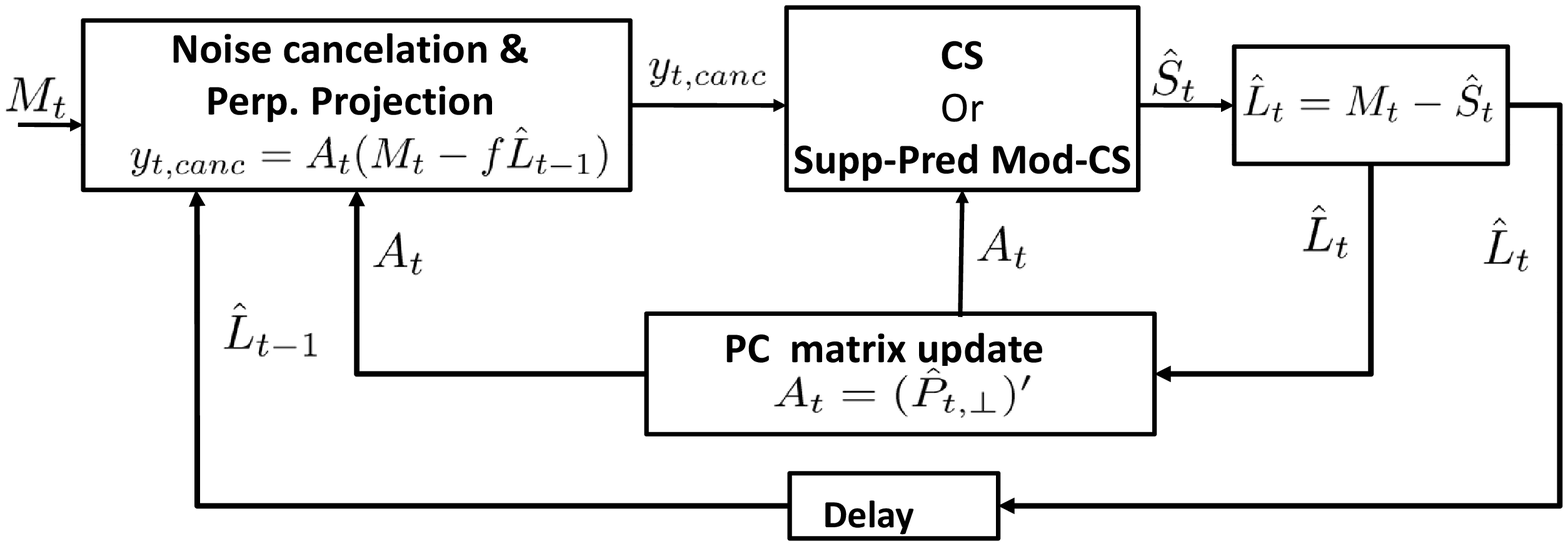, width = 8cm, height=3.5cm}\label{chart1}
}
\subfigure[Support-Predicted Modified-CS]{
\psfig{file = 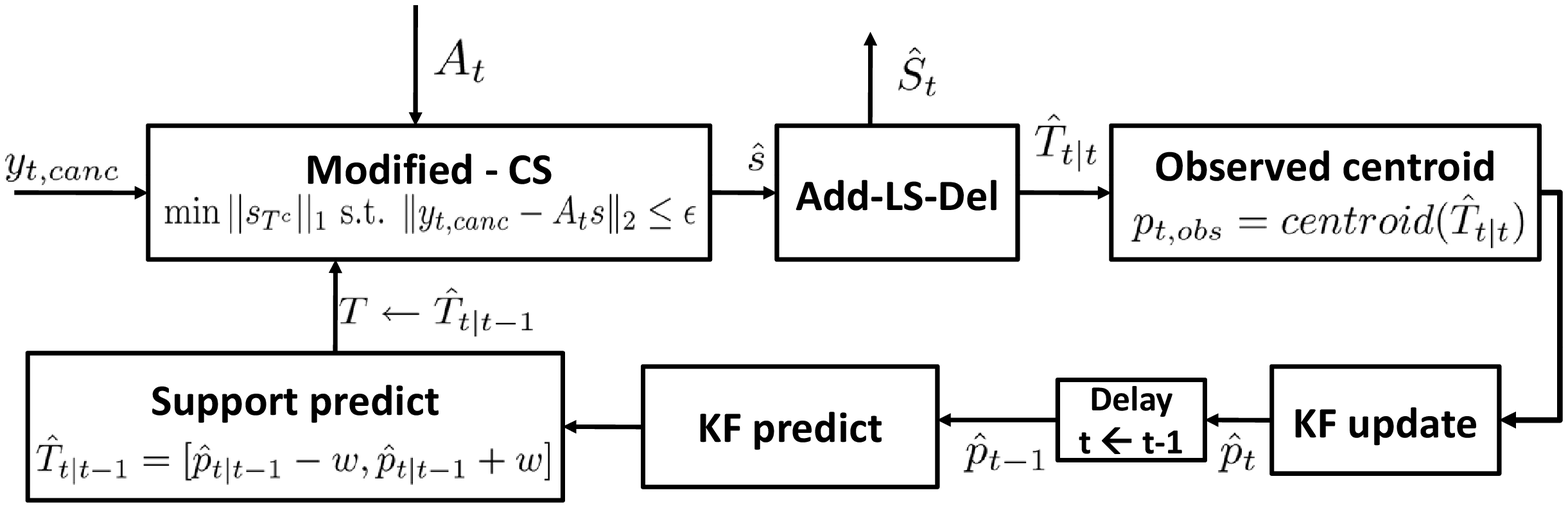, width = 10cm, height=4cm}\label{chart2}
}
}
\vspace{-0.1in}
\caption{Block Diagram of SuppPred-ModCS-RRPCP}\label{complete_diag}
\vspace{-0.2in}
\end{figure*}

The key idea of Recursive Robust PCP (RRPCP) \cite{rrpcp_allerton} is as follows. Assume that the current PC matrix $P_t$ has been estimated. We project the outlier-corrupted data vector, $M_t$, into the space perpendicular to it, i.e. we compute
$$y_t:=A_t M_t, \ \ \text{where} \ \ A_t:= (\hat{P}_{t,\perp})'$$
and $\hat{P}_{t,\perp}$ is \emph{an} orthogonal complement of $\hat{P}_t$. If $M_t$ is $n$ dimensional, then the dimension of $y_t$ is $n-r$ where $r:=rank(\hat{P}_t)$.
Notice that
$$y_t = A_t S_t + \beta_t, \  \ \beta_t:= A_t L_t = A_t P_t (x_t)_{N_t}.$$
If $\hat{P}_t \approx P_t$, then $A_t P_t \approx 0$, i.e. this should nullify most of the contribution of the low rank part, so that $\beta_t$ can be interpreted as small ``noise".
Finding the sparse outlier, $S_t$, from this projected data vector, $y_t$, now becomes a noisy sparse reconstruction / compressive sensing (CS)  \cite{bpdn,candes,donoho} problem. We can solve
\begin{equation}
\min_s \|s\|_1 \   \text{subject to}  \ \|y_t - A_t s\|_2 \le \epsilon
\label{rpcp_simple_eq}
\end{equation}
with $\epsilon$ chosen as explained in \cite{rrpcp_allerton}. Denote its output by $\hat{S}_t$. We can then estimate $\hat{L}_t = M_t - \hat{S}_t$ which can be used to recursively update $\hat{P}_t$ every-so-often as in \cite[Algorithm 1]{rrpcp_allerton}. When a new direction gets added to $P_t$, but $\hat{P}_t$ is not yet updated, the ``noise", $\beta_t$, will start increasing. We explain how to ``cancel" some of it in Sec. \ref{ModelL} after explaining the model on $L_t$. The block diagram is shown in Fig. \ref{chart1}.
Our idea is also somewhat related to that of \cite{decodinglp,rpca_regression} in that all of these also try to cancel the ``low rank" part by projecting the original data vector into the perpendicular space of the tall matrix that spans the ``low rank" part. But the big difference is that in all these, this matrix is \emph{known}. In our problem  this matrix (the PC matrix) is \emph{unknown and can change with time.}%

Now, if the support size of the sparse outliers increases for a given rank $r$, the number of projected ``measurements" available for the CS step, $n-r$, may become too small for CS to work.
But notice that the correlated sparse outliers (e.g. moving foreground objects in video) can be interpreted as sparse signal sequences \emph{whose support change over time is either very slow; or quite often is not slow, but is still highly correlated, e.g. the support can ``move" or ``expand" or change according to some model, over time.} By using a model on how the objects move/deform, or other models on how the support changes, it should be possible to obtain a support prediction that can serve as an accurate ``partial support knowledge". We can then tap into our recent work on Modified-CS which solves the sparse recovery problem when partial support knowledge is available \cite{modcs_ts}. Denote the partial support knowledge by $T$. Modified-CS tries to find the solution that is sparsest outside the set $T$ among all solutions that satisfy the data constraint. It does this by replacing $ \|s\|_1 $ in (\ref{rpcp_simple_eq}) by $ \|s_{T^c}\|_1 $.
As proved and experimentally shown in \cite{stability_allerton}, in the noisy case, it has stable and small error using much fewer measurements than what simple CS needs.

In this work, we demonstrate the above idea by assuming that the foreground contains a single \emph{rigid} moving object that follows a constant velocity model (with small random acceleration). We use a Kalman filter (KF) to track its motion over time.
The KF predicted location of the object and its size tell us its predicted support at the current time. This is then used to solve Modified-CS and obtain an updated support estimate. The centroid (or median) of this support set tells us the observed location of the object, which may be erroneous because our support estimate is not perfect. This serves as the noisy observation for the KF update step.%

\Subsection{Notation}
The set operations $\cup$, $\cap$ and $\setminus$ have the usual meanings. For any set $T \subset \{1 , \cdots n\}$, $T^c$ denotes its complement and $|T|$ denotes its cardinality.

For a matrix $A$, we let $A_i$ denote the $i^{th}$ column of $A$ and we let $A_T$ denote a matrix composed of the columns of $A$ indexed by $T$. We use $A'$ to denote transpose, and $A^{\dag}$ to denote its pseudoinverse. For a tall matrix $A$, $A^{\dag} = (A'A)^{-1}A'$.

For vector $v$, $v_i$ denotes the $i$th entry of $v$ and $v_T$ denotes a vector consisting of the entries of $v$ indexed by $T$. We $\|v\|_k$ to denote the $\ell_k$ norm of $v$. The support of $v$, $\text{supp}(v)$, is the set of indices at which $v$ has nonzero value, $\text{supp}(v): = \{i:\ v_i \neq 0\}$.

\Subsection{Paper Organization}
We define the problem and give the correlation models in Sec. \ref{models}. 
Our proposed method, Support-Predicted Modified-CS RRPCP (SuppPred-ModCS-RRPCP), is described in Sec. \ref{supppred_sec}, where we also argue why it should be stable. Experiments showing that SuppPred-ModCS-RRPCP provide a significant gain than RRPCP \cite{rrpcp_allerton} and PCP \cite{rpca} are given in Sec. \ref{expt_sec}.

\Section{Problem Definition and Correlated Models}\label{models}
We give the problem definition below and explain the correlation models on $L_t$ and $S_t$ in Sec. \ref{ModelL} and \ref{ModelS} respectively.

\Subsection{Problem Definition}
Consider the problem of tracking moving foreground objects (correlated sparse outliers) in a slowly changing background (low rank part).
For explaining our ideas in a simple fashion, we assume a 1D ``image" with one moving foreground object. The extension to 2D image and multiple moving objects is explained in Sec. \ref{expt_sec}. Our simulations use this case.

In video applications, the outlier (moving object) is not added, but is overlaid. In other words, at each $t$, the image $M_t$ satisfies
\begin{equation}
(M_t)_i = \left\{ \begin{array}{ll} \label{overlay}
(O_t)_i  & \mbox{\  \ $i \in T_t$} \\
(L_t)_i  & \mbox{\  \ $i \in T_t^c$}
\end{array}
\right.
\end{equation}
where $O_t$ is the sparse foreground image and $T_t$ is its support, i.e. $(O_t)_{T_t^c} = 0$.
and $L_t$ is the background image.
However, this problem can be rewritten in the form (\ref{eq1}) by letting 
\begin{equation}
(S_t)_i = \left\{ \begin{array}{ll}
(O_t-L_t)_i \  & \ \mbox{ $i \in T_t$}  \label{S_def}\\
 0           \ &  \ \mbox{ $i \in T_t^c$}
\end{array}
\right.
\end{equation}
Notice that $S_t$ and $O_t$ have the same support $T_t$.

\Subsection{Model on $L_t$ (background image)} \label{ModelL}
For $L_t$, we use the model from \cite{rrpcp_allerton}. Briefly, we assume that the support of $x_t$, $N_t$, is piecewise constant with time, and also assume that each nonzero element of $x_t$ is
independent and follows a piecewise stationary model with nonstationary transients between pieces.
Specifically, an independent first order autoregressive (AR) model is assumed for each nonzero element of $x_t$,
all with the same AR parameter $0<f<1$ as given in \cite{rrpcp_allerton}.
Every $d$ frames, there are some indices get added or deleted from $N_t$.
When a new direction added to $P_t$, initially $x_t$ along that direction starts with some initial small value,
but slowly increases to a stable large value.
Before an existing direction gets deleted from $P_t$, $x_t$ along that direction decays exponentially to zero.

When new directions are added to $P_t$, but are not part of the current $\hat{P}_t$, the ``noise", $\beta_t$, will start to increase. But by using the AR model, we can cancel some of the noise by replacing $y_t$ with $y_{t,canc}$ defined as
\begin{equation}
y_{t,canc} := y_t - f A_t \hat{L}_{t-1} \label{ycancel}
\end{equation}
Now, the ``noise" is only $\beta_t - f\hat{\beta}_{t-1}$ which is much smaller than $\beta_t$ if $f$ is close to 1 and if $\hat{L}_{t-1} \approx L_{t-1}$. 

\Subsection{Model on Support Change of $S_t$ (foreground image)} \label{ModelS}
Let $p_t$ be the location of the foreground object's centroid, let $v_t$ denote its velocity and let $w$ denote its width. Thus, its support is,
$$T_t = [p_t-w, p_t +w ]$$
Let
\begin{equation*}
g_t := \left[ \begin{array}{c}p_t\\ v_t \\ \end{array} \right] \ \text{and} \
F := \left[ \begin{array}{cc} 1 & \ 1 \\ 0 & \ 1 \\ \end{array}\right].
\end{equation*}
We assume a constant velocity model with small random acceleration \cite{poor_book} on the object's motion, i.e.,
\begin{equation}
g_t = F g_{t-1} + n_t \ \    \label{obsmod3}
\end{equation}
where $n_t$ is bounded noise with zero mean and variance $Q$. The variance matrix $Q$ is of the form  $\left[
                                                                \begin{array}{cc}
                                                                  0 \ & 0 \\
                                                                  0 \ & q \\
                                                                \end{array}
                                                              \right]$
because we only add noise to velocity and not to position \cite{poor_book}.

\Section{Support-Predicted Modified-CS Recursive Robust Principal Components' Pursuit}\label{supppred_sec}
In this section, we explain our algorithm. 

\Subsection{Support-Predicted Modified-CS}
Recall that in RRPCP, the number of projected ``measurements" available for the CS step is $n-r$ where $r=rank(\hat{P}_t)$.
If the support size of the sparse part, $|T_t|$, increases for a given $r$, then $n-r$ may become too small for CS to work. In this section, we show how to use the correlated support change of $S_t$ along with Modified-CS to address this problem. The overall idea is as follows. The support of $S_t$ follows the  model given Sec. \ref{ModelS}. We use this in a KF to obtain its location prediction, which can then give us its support prediction $\hat{T}_{t|t-1}$. We then solve Modified-CS, given in (\ref{rpcp_modCS_eq}), with $T = \hat{T}_{t|t-1}$. The support estimate of the Modified-CS output serves as the updated support, $\hat{T}_{t|t}$, and the centroid\footnote{One can replace the centroid by the median for more robustness.} of this updated support serves as the ``observed" location, $p_{t,obs}$, for the KF update step to update the object's location estimate. We explain each of these steps below.

\begin{algorithm}[th!]
\caption{Support-Predicted Modified-CS}\label{algo1}
\begin{itemize}
\item [1)] Predict centroid by (\ref{KFpred1}) and (\ref{KFpred2})
\item [2)] Predict support by (\ref{Tpred})
\item [3)] Update support
\begin{itemize}
\item Modified-CS: solve (\ref{rpcp_modCS_eq}) using $T = \hat{T}_t$.
\item Add-LS-Del procedure:
\begin{eqnarray}
T_{add} &=& T \cup \{i \in T^c: |(\hat{s})_i| > \alpha_{add} \} \label{ald1} \\
(\hat{s})_{T_{add}} &=& ((A_t)_{T_{add}})^{\dag} y_{t,canc}, \ (\hat{s})_{T_{add}^c} = 0 \label{ald2} \\
\hat{T}_{t|t} &=& T_{add} \setminus  \{i \in T_{add} : |(\hat{s})_i| < \alpha_{del} \} \label{ald3}\\
(\hat{S}_t)_{\hat{T}_{t|t}} &=& ((A_t)_{\hat{T}_{t|t}})^{\dag} y_{t,canc}, \ (\hat{S}_t)_{\hat{T}_{t|t}^c} = 0 \label{ald4}
\end{eqnarray}
\end{itemize}
\item [4)] Update centroid by (\ref{KFupd1}), (\ref{KFupd2}) and (\ref{KFupd3}).
\end{itemize}
\end{algorithm}

\emph{Predict Centroid:}
Let $\hat{g}_{t|z} = [\hat{p}_{t|z} \ \hat{v}_{t|z}]'$ represent the estimate of $g_t$ at time $t$ given measurements up to, and including at time $z$. Similar rule applies for $\hat{T}_{t|z}$.
Let $\Sigma_{t|t-1}$, $\Sigma_{t|t}$ and $K_t$ denote the prediction and updated error covariance matrices and the Kalman gain used by the KF. Compute
\begin{eqnarray}
\hat{g}_{t|t-1} &=& F \  \hat{g}_{t-1|t-1} \label{KFpred1} \\
\Sigma_{t|t-1} &=& F \ \Sigma_{t-1|t-1} \ F' + Q \label{KFpred2}
\end{eqnarray}

\noindent \emph{Predict Support:}
We can get a reliable support prediction as
\begin{equation}
\hat{T}_{t|t-1} = [\hat{p}_{t|t-1} - w, \hat{p}_{t|t-1} + w] \label{Tpred}
\end{equation}

\noindent \emph{Update Support using Modified-CS:}
Assuming $\hat{T}_{t|t-1}$ is a good support prediction, we can use it as the partial support knowledge for Modified-CS, i.e. we can solve
\begin{equation}
\min_s \|s_{T^c}\|_1  \ \ \text{subject to} \ \ \|y_{t,canc}  - A_t s \|^2 \le \epsilon
\label{rpcp_modCS_eq}
\end{equation}
with $T = \hat{T}_{t|t-1}$.
Let $\hat{s}$ be the solution of (\ref{rpcp_modCS_eq}) with $T = \hat{T}_{t|t-1}$. As explained in \cite{stability_allerton}, since $\hat{s}$ is biased towards zero along $T^c$ and it may be biased away from zero along $T$ (there is no cost or constraint on $s_T$), we will run into problems if we try to use a single threshold for support estimation. A better approach is the use the Add-LS-Del procedure as summarized in step 3 of Algorithm \ref{algo1}. This was first introduced in our older work \cite{LS-CS-residual,kfcsicip} and simultaneously also in \cite{SubSpaceCS,cosamp}. It involves a support addition step (that uses a smaller threshold), as in (\ref{ald1}), followed by LS estimation on the new support estimate, $T_{add}$, as in (\ref{ald2}), and then a deletion step that thresholds the LS estimate, as in (\ref{ald3}). This can be followed by a second LS estimation using the final support estimate, as in (\ref{ald4}). The addition step threshold, $\alpha_{add}$, needs to be just large enough to ensure that the matrix used for LS estimation, $A_{T_{add}}$ is well-conditioned. If $\alpha_{add}$ is chosen properly, the LS estimate on $T_{add}$ will have smaller error than the Modified-CS output. As a result, deletion will be more accurate when done using this estimate. This means that one can also use a larger $\alpha_{del}$ to ensure quicker deletion of extras.

\emph{Update Centroid:}
Let $p_{t,obs}$ denote the ``observed" centroid of foreground object at time $t$ obtained by
taking the mean of the updated support estimate $\hat{T}_{t|t}$,
i.e., $p_{t,obs} = centroid(\hat{T}_{t|t})$. Our observation model is
\begin{equation}
p_{t,obs} = H g_t + \omega_t   \label{obsmod4}
\end{equation}
where $H : = [ 1 \ 0]$ and $\omega_t$ is observation error, which is assumed to be zero mean with variance $R$. This arises because there are extras and misses in $\hat{T}_{t|t}$ and hence $p_{t,obs} = centroid(\hat{T}_{t|t}) \neq centroid(T_t)=p_t$. In this work, $R$ is heuristically selected, but, in general, one can approximate it using simplifying assumptions on the support computation.%

The KF update step is as follows.
\begin{eqnarray}
K_t &=& \Sigma_{t|t-1} \ H' \ (H\ \Sigma_{t|t-1} \ H'+R)^{-1} \label{KFupd1} \\
\hat{g}_{t|t} &=& \hat{g}_{t|t-1} + K_t \ (p_{t,obs}-H\ \hat{g}_{t|t-1}) \label{KFupd2}\\
\Sigma_{t|t} &=& \Sigma_{t|t-1} - K_t \ H \ \Sigma_{t|t-1} \label{KFupd3}
\end{eqnarray}

The above steps for Support-Predicted Modified-CS are summarized in Algorithm \ref{algo1} and in block diagram of Fig. \ref{chart2}.%

\Subsection{Complete Algorithm of SuppPred-ModCS-RRPCP}
With the support estimate $\hat{T}_{t|t}$ and sparse estimate $\hat{S}_t$ obtained in Algorithm \ref{algo1}, $O_t$ and $L_t$ can be estimated as
\begin{eqnarray}
(O_t)_{\hat{T}_{t|t}} &=& (M_t)_{\hat{T}_{t|t}}, \ (O_t)_{\hat{T}_{t|t}^c} = 0 \label{fb1} \\
\hat{L}_t &=& M_t - \hat{S}_t \label{fb2}
\end{eqnarray}
Also, $\hat{P}_t$ can be updated as in \cite[Algorithm 1]{rrpcp_allerton}.

A complete algorithm incorporating the idea of Support-Predict Modified-CS is given in Algorithm \ref{algo2}.
\begin{algorithm}[t!]
\caption{SuppPred-ModCS-RRPCP}
\label{algo2}
At $t=t_0$, suppose a good estimate of PC matrix, $\hat{P}_0$ is available from training data.
%
For $t>t_0$, do the following:
\begin{itemize}
\item [1)] Update $\hat{P}_t$ using Algorithm 1 in \cite{rrpcp_allerton} which is based on \cite{sequentialSVD} and correspondingly, update $A_t := (\hat{P}_{t,\perp})'$.
\item [2)] Obtain $y_{t,canc}$ by (\ref{ycancel}).
\item [3)] Support-Predict Modified-CS using Algorithm \ref{algo1}.
\item [4)] Estimate $\hat{O}_t$ and $\hat{L}_t$ by (\ref{fb1}) and (\ref{fb2}).
\item [5)] Increment $t$ by $1$ and go to step 1).
\end{itemize}
\end{algorithm}

For simplicity, we have presented SuppPred-ModCS-RRPCP for a 1D image with one moving foreground object.
However, it can be easily extended to the 2D case with multiple moving objects. We explain how to do it in Sec. \ref{expt_sec}

\Subsection{Discussion}\label{dissec}
SuppPred-ModCS-RRPCP is a recursive approach.
Hence an important question is when and why will it be stable?
We try to give an induction argument here that we will formalize in later work. The key idea is as follows. Everywhere in this discussion, ``bounded" means bounded by a time-invariant value. Suppose that at $t-1$, $|p_{t-1} - \hat{p}_{t-1|t-1}|$ is bounded and small. Since $n_t$ is bounded and small, this means that $|p_{t} - \hat{p}_{t|t-1}|$ is also bounded and small. This, in turn will mean that the same holds for the support prediction errors $|T_t \setminus \hat{T}_{t|t-1}|$ and $|\hat{T}_{t|t-1} \setminus T_t|$. Using this and arguments similar to those in \cite{stability_allerton}, the support update step will also result in $\hat{T}_{t|t}$ with bounded and small extras and misses (in fact the bound on extras is zero). This step will require showing that the ``noise" seen by modified-CS, $\beta_t - f A_t \hat{L}_{t-1}$, is bounded; that most nonzero elements of $S_t$ are large enough; and that $A_t$ satisfies certain conditions. Finally, since $\hat{T}_{t|t}$ has zero extras, we will just need to argue that the misses, $T_t \setminus \hat{T}_{t|t}$, will result in bounded and small centroid observation error, $\omega_t$. 
This will finally ensure bounded and small $|p_t - \hat{p}_{t|t}|$. This, along with ensuring stability of $\Sigma_{t|t}$, will ensure bounded and small  $|p_t - \hat{p}_{t|t}|$.  Our simulations given in Sec. \ref{expt_sec} do indicate that SuppPred-ModCS-RRPCP is stable.

We use the KF in this paper, but in general the above argument will go through with any stable linear observer.

\begin{figure}
\centerline{
\psfig{file = 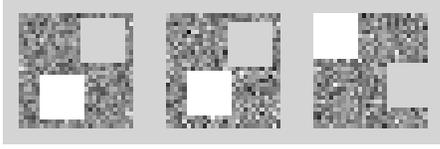,height=2cm}
}
\vspace{-0.07in}
\caption{Images at time $t=1,5,100$.}\label{org}
\vspace{-0.2in}
\end{figure}

\Section{Numerical Experiments}\label{expt_sec}
We evaluate the performance of SuppPred-ModCS-RRPCP in the 2D case for the problem of tracking
two moving objects in a simulated image sequence of size $28 \times 28 \times 100$. Fig.\ref{org} shows the image frame at $t=1,5,100$. One image written as a 1D vector, $M_t$, is $n$ dimensional with $n=28^2=784$ and it satisfies (\ref{overlay}).

The background image, $L_t $ is simulated according to the model in Sec.\ref{ModelL}. Initially, there are $350$ principal directions in the PC matrix $P_t$. At $t=5$, one new direction is add to $P_t$ with a small variance and it slowly stabilizes. At $t=30$, one existing direction starts to decay exponentially  to zero. Thus, $r:=rank(\hat{P}_t) \approx 350$.

The foreground overlay, $O_t$, consists of two $11 \times 11$ blocks that have different constant intensity $80$ and $50$.
These two objects move independently. Each object moves along horizonal and vertical directions independently with some initial location and velocity satisfying (\ref{obsmod3}). In (\ref{obsmod3}) and (\ref{obsmod4}), we use zero mean truncated gaussian noise with variance $q=10^{-4}$
and $R = 10^{-3}$ for $n_t$ and $\omega_t$.
Note that $|T_t| : = |supp(S_t)| \approx 242$, while $n-r$ is only about $434$.

Fig. \ref{plot1} shows the reconstruction error of $S_t$ for two online methods, Algorithm \ref{algo2} and RRPCP \cite{rrpcp_allerton}, as well as offline PCP \cite{rpca}. For offline PCP, we use the entire sequence and show the error for each image frame separately.
PCP fails due to the correlation of $S_t$. The other two online methods work well because they do not require the support and signs of $S_t$ to be i.i.d. Algorithm \ref{algo2} outperforms RRPCP greatly because it utilizes the correlation model of $S_t$ while RRPCP does not.

With an estimate of $T_t$, we separate different objects by thresholding the intensity of $M_t$ on the support estimate. This is needed for running two separate KFs for each of their centroids. 

\begin{figure}
\centerline{
\psfig{file = 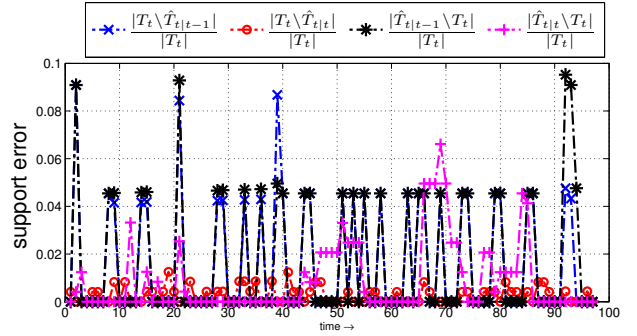, height = 4.5cm}
}
\vspace{-0.1in}
\caption{Support error of Support-Predicted Modified-CS}
\vspace{-0.25in}
\label{plot2}
\end{figure}

We show the number of extras and misses in $\hat{T}_{t|t-1}$ 
and $\hat{T}_{t|t}$ 
 when using Algorithm \ref{algo2} in Fig. \ref{plot2}. Recall that $\hat{T}_{t|t-1}$ is the predicted support used by Modfied-CS (\ref{rpcp_modCS_eq}), and $\hat{T}_{t|t}$ is the updated support estimate obtained by Modfied-CS followed by Add-LS-Del (\ref{ald1})-(\ref{ald4}). Clearly, this corrects a lot of the prediction errors.

\bibliographystyle{ieeepes}
\bibliography{tipnewpfmt}

%

\end{document}